\documentclass[prl,twocolumn,groupedaddress]{revtex4-1}
\usepackage{graphicx}% Include figure files
\usepackage{dcolumn}% Align table columns on decimal point
\usepackage{bm}% bold math

\usepackage{amsmath}
\usepackage{graphicx}
\usepackage{color}
\usepackage{wrapfig}
\usepackage{epstopdf}

\graphicspath{{pics/}}

\newcommand{\om}{\omega}
\newcommand{\bee}{\begin{equation}}
\newcommand{\ene}{\end{equation}}

\newcommand{\bea}{\begin{eqnarray}}
\newcommand{\ena}{\end{eqnarray}}

\begin{document}

\title{Demonstration of the enhanced Purcell factor in all-dielectric structures}

\author{Alexander Krasnok,$^{1}$ Stanislav Glybovski,$^{1}$ Mihail Petrov,$^{1}$ Sergey Makarov,$^{1}$\\ Roman Savelev,$^{1}$ Pavel Belov$^{1}$, Constantin Simovski$^{1,2}$, and Yuri Kivshar$^{1,3}$}
\address{
$^{1}$ITMO University, St.~Petersburg 197101, Russia\\
$^{2}$Aalto University, School of Electrical Engineering
P.O. Box 13000, 00076 Aalto, Finland\\
$^{3}$Nonlinear Physics Center, Australian National University, Canberra ACT 2601, Australia}

\begin{abstract}
The Purcell effect is usually described as a modification of the spontaneous decay rate in the presence of a resonator. In plasmonics, this effect is commonly associated with a large local-field enhancement in "hot spots" due to the excitation of surface plasmons. However, high-index dielectric nanostructures, which become the basis of all-dielectric nanophotonics, can not provide high values of the local-field enhancement due to larger radiation losses. Here, we demonstrate how to achieve a strong Purcell effect in all-dielectric nanostructures, and show theoretically that the Purcell factor can be increased by two orders of magnitude in a finite chain of silicon nanoparticles. Using the eigenmode analysis for an infinite chain, we demonstrate that the high Purcell factor regime is associated with a Van Hove singularity. We perform a proof-of-concept experiment for microwave frequencies and observe the 65-fold enhancement of the Purcell factor in a chain of 10 dielectric particles.
\end{abstract}

\maketitle

All-dielectric nanophotonics based on nanoparticles of highly refractive materials allows to control not only electric but also magnetic component of light at the nanoscale without high dissipative losses, inherent for metallic (plasmonic) nanostructures~\cite{Cummer_08, Evlyukhin:NL:2012, Miroshnichenko:NL:2012, Brener_12, Lukyanchuk13, Krishnamurthy_13}. The recent progress in this ''magnetic light'' concept has opened a door to many important applications of such structures including metamaterials~\cite{Cummer_08, Brener_12}, metasurfaces~\cite{Staude_15}, sensors and nanoantennas~\cite{Miroshnichenko:NL:2012, Maier_2015_NC}. However, the development of quantum optics and biological sensors based on all-dielectric nanostructures requires high degree of their interaction with quantum emitters (molecules, quantum dots, defects in solids), or in other words strong Purcell effect. This effect is usually defined as a modification of the spontaneous emission rate of a quantum emitter induced by its interaction with environment~\cite{Purcell_46,Sauvan2013a, Krasnok_Purcell_2015}. Although the Purcell effect was discovered in the context of nuclear magnetic resonance~\cite{Purcell_46}, nowadays it is widely used in many applications, ranging from microcavity light-emitting devices~\cite{Fainman_2010} to single-molecule optical microscopy~\cite{Koenderink_PRL_11, Cosa_2013}, being also employed for tailoring optical nonlinearities~\cite{Soljacic_2007}, and enhancing spontaneous emission from quantum cascades~\cite{Minot_2007}.

It is generally believed that large values of the Purcell factor are observed in the systems with strong local field enhancement associated with the formation of "hot spots" (e.g. in plasmonic nanoantennas)~\cite{Koenderink2010, Sauvan2013a}. Accordingly, in order to achieve high values of the Purcell factor, a quantum emitter should be placed in one of such hot spots. At the same time, in contrast to their plasmonic counterparts, all-dielectric nanostructures do not demonstrate strong electric field enhancement~\cite{Kuznetsov15}, which is believed to be the main reason of small values of the Purcell factor. This originates from large radiation losses and preferential localization of light energy inside the dielectric nanostructures rather than at the surfaces. Indeed, by comparing the studies of the Purcell effect in dielectric and plasmonic nanostructures, one can conclude that dielectric structures demonstrate the Purcell factor of one-two orders of magnitude smaller even in systems of several particles~\cite{Schmidt_12, KrasnokOE, Aizpurua_2013, Krasnok_Purcell_2015, Belov_OS_15}. Namely, the Purcell factor does not exceed 10-15 in all previously published papers on dielectric nanostructures with strong dipole magnetic response~\cite{Schmidt_12, KrasnokOE, Schuller:OE:2009, Bonod2015, Krasnok_Purcell_2015}. Therefore, such small values of the Purcell factor do not allow to achieve an effective interaction of nanostructures with light emitters.

\begin{figure}[!b]
\includegraphics[width=0.5\textwidth]{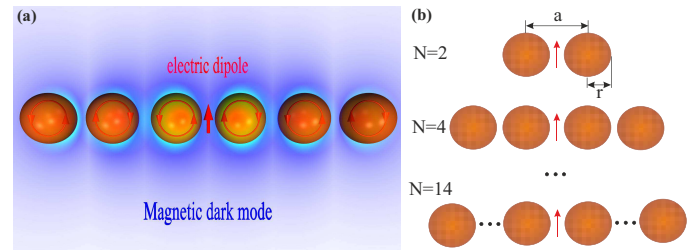}
\caption{(a) Excitation of a dark magnetic mode in the chain of dielectric nanoparticles by an electrical dipole emitter. (b) All-dielectric dimer (N=2), quadrumer (N=4), ..., and dekatesseramer (N=14) nanoantennas in the form of chains with the same period a$=$200~nm. The dielectric constant of the nanoparticles is $\varepsilon=16$, the radii of nanoparticles are r=70~nm.}\label{esciz}
\end{figure}

\begin{figure*}
\includegraphics[width=0.99\textwidth]{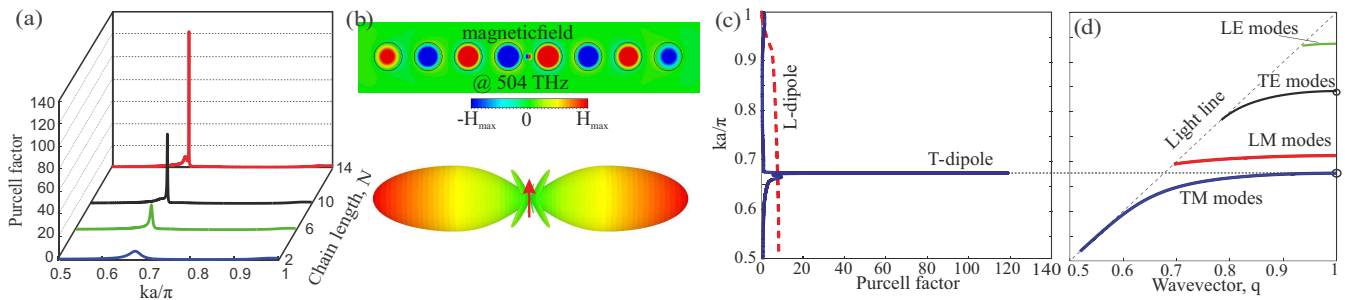}
\caption{(a) The dependences of Purcell factor for different chain length N=2-14 for T--dipole orientation on radiation frequency. (b) Top: the magnetic field distribution at 504~THz for the case of N=8. Bottom: 3D radiation pattern at 504~THz for the same structure. (c) Spectrum of Purcell enhancement computed for $T$ and $L$ polarizations of the dipole emitter for finite chain consisting of N$=$14 nanoparticles. (d) Dispersion curves of eigenmodes in infinite dielectric chain.}\label{Fig2_theor}
\end{figure*}

In this Letter, we reveal how to achieve the strong Purcell effect for all-dielectric nanostructures without high local-field enhancement. This becomes possible due to excitation of a specific dark mode at the band edge of a nanoparticles chain [as shown schematically in the Fig.~\ref{esciz}(a)], which represent a special type of modes which can not be excited by a plane wave~\cite{magdark_1, magdark_2, magdark_3, magdark_4, magdark_5, Yang2015}. When nanostructure is radiated by a dipole light source, these modes are excited contributing to the local density of states. As an example, in the recent study~\cite{KrasnokNanoscale}, superdirectivity of all--dielectric nanoantennas was achieved in a similar way. Nearly zero group velocity at the band edge results in a \textit{Van Hove singularity}, manifesting itself in dramatic enhancement of local density of states. Thus, even if the local-field enhancement of a nanostructure is not strong, the high Purcell factor can be achieved through the special arrangement of the geometry and dipole light source. Based on this, here we show that the Purcell factor can be increased by two orders of magnitude in a finite chain of crystalline silicon (c-Si) nanoparticles, demonstrating previously much lower values. Using an eigenmode analysis for an infinite chain, we reveal that the high Purcell factor is associated with the Van Hove singularities. Finally, we perform a proof-of-concept experiment for microwave frequencies and observe the Purcell factor enhancement up to 65 times for a chain of 10 particles. Our approach would allow to increase the Purcell factor even more by increasing the number of dielectric particles in the system.

We start with the simplest system of an all-dielectric dimer consisting of two dielectric nanospheres with a high refractive index (such as a crystalline silicon). For analysis we choose a dielectric material with the dielectric constant of $\varepsilon=16$. The nanosphere of this material with radius of $r=70$~nm has the magnetic dipole resonance at a frequency of 500~THz. The distance between the centers of the nanoparticles is equal to $a=200$~nm. We have placed an electric dipole emitter exactly in the middle between the nanoparticles orthogonally their axial axis [as presented in Fig.~\ref{esciz}(b)]. In this case, at the magnetic resonance frequency of the nanoparticles, their magnetic moments oscillate with a phase difference $\pi$, i.e. in antiphase. We calculate the Purcell factor (F) by using the well-known formula~\cite{Novotny_Hecht_book}:
\begin{equation}\label{eq0}
{\rm F}=1+\frac{6\pi\varepsilon_0}{k^{3}|{\bf d}|^{2}}\mbox{Im}\left[ {\bf d}^{*}\cdot{\bf E}_s({\bf R}_0)\right],
\end{equation}
where $k$ is the wave number, $\mathbf{d}$ is the radiating electric dipole moment of an emitter, and $\mathbf{E}_s(\mathbf{R}_0)$ is a scattered electric field at the emitter origin $\mathbf{R}_0$ produced by the nanoantenna, $\varepsilon_0$ is the vacuum permittivity. The Purcell factor of the all-dielectric dimer reaches 10 (see Fig.~\ref{Fig2_theor}(a), blue curve). The frequency at which the maximum value of the Purcell factor of this dimer nanoantenna is observed is close to the frequency of the magnetic resonance of a single nanosphere.

Consequent doubling of the number of nanospheres $N$ [see Fig.~\ref{esciz}(b)], we observe a significant increase of Purcell factor from 10 ($N=2$) up to 120 ($N=14$). The Purcell factor of this nanoantennas have been also calculated using the Green's function approach (\ref{eq0}). The results are presented in the Fig.~\ref{Fig2_theor}(a), where the dimensionless frequency $\omega=ka/\pi$ is used. In the Fig.~\ref{Fig2_theor}(b) the magnetic field distribution for the case of N=8 (top) and 3D radiation pattern for the same structure (bottom) at the frequency of 504~THz are presented. These results show that at the maximum of Purcell factor the magnetic dipoles of each pair of nanoparticles oscillate in opposite phase. In this regime the emitted light is localized to the chain and the radiation pattern has two narrow lobes directed along the chain. For example, in the Fig.~\ref{Fig2_theor}(c) the Purcell factor dependences on the frequency for the case $N=14$ for parallel (L--dipole, dashed curve) and perpendicular (T--dipole, solid curve) orientations of the dipole emitter are presented. The maximum value of the Purcell factor increases at the frequency 504~THz where each nanoparticle is polarized in antiphase in relation to its neighboring nanoparticles.

\begin{figure*}[!t]
\includegraphics[width=0.99\textwidth]{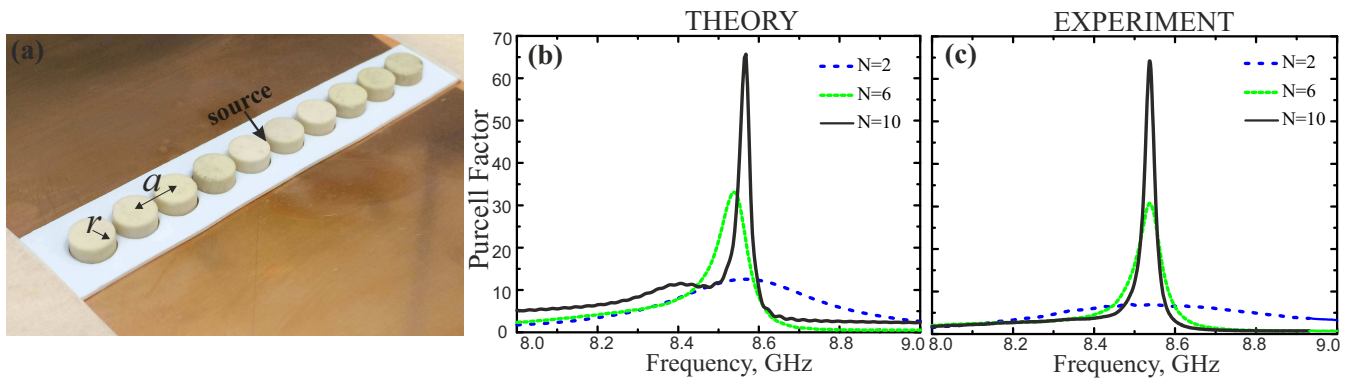}
\caption{Experimental verification of the Purcell enhancement. (a) The geometry of the experimental setup. (b) Numerical simulations of the Purcell factor for a chain of $N=2-10$ dielectric disks. (c) Experimentally measured Purcell factor by 3-mm dipole antenna placed in the center of the chain. The inset shows the geometry. Parameters of numerical simulations and experiment are: $r=4$~mm, $H=4$~mm, $a=5$~mm, and $\varepsilon=16$.}
\label{Fig3_exp}
\end{figure*}

Next, to clarify the nature of the very narrow Purcell factor resonance, we consider the case of an infinite chain. The dispersion properties of an infinite chain with period $a=200$~nm consisting of dielectric nanospheres with permittivity $\varepsilon=16$ and radius $r=70$~nm embedded in free space are illustrated by Fig.~\ref{Fig2_theor}(d). Here the dipole-dipole interaction model~\cite{Weber2004} has been used: each sphere is replaced by its electric and magnetic dipoles. The set of eigenmodes consists of four branches: transverse electric (TE), transverse magnetic (TM), longitudinal electric (LE), and longitudinal magnetic (LM). The corresponding dispersion curves $\om(q)$ are shown in Fig.~\ref{Fig2_theor}(d), where the dimensionless quasi vector $q=\beta a/\pi$ is used, where $\beta$ is the propagation constant. The leaky branches lying above the light line are not shown. The Purcell factor spectra for a dipole polarized along (L--dipole) and perpendicular (T--dipole) to the chain axis are shown in Fig.~\ref{Fig2_theor}(c) (for $N=14$) for ease of comparison with the eigenmodes.
One can see that the multifold enhancement of the Purcell factor is predicted for T--dipole at the frequency corresponding to TM--mode band edge. Having almost zero group velocity at the band edge the chain modes has divergent density of states due to \textit{Van Hove singularity} in the infinite 1D structure, that may result in ultrahigh Purcell factor~\cite{Fink_2004}. However, the high Purcell factor is observed for TM--mode only, and is not observed for L--polarized dipole in the whole spectrum. The explanation of this effect lies in the symmetry of the band edge modes. The phase shift between the dipole moments of the neighboring nanoparticles equals to $\pi$ at the band edge that makes these modes to be sufficiently ''dark''. The coupling between the dipole emitter and TE--, LE-- and LM--modes is suppressed due to the \textit{symmetry mismatch}. It can be illustrated on the example of T--dipole and TE--mode: the electric fields induced by the oppositely polarized nanoparticles cancel each other in the point of the dipole emitter. Thus, the excitation of TE--mode by T-dipole is principally weak that also applies to LE-- and LM--modes. However, the TM--mode possesses the required symmetry and the electric fields generated by neighboring nanoparticles compliment each other, thus, the TM--mode is effectively excited. Having high density of states these modes give rise to the Purcell factor. Figure~\ref{Fig2_theor}(b) shows that this effect relates to collective excitation of dark mode. One can see that Purcell factor rapidly increases with the increase of chain length that is explained by forming mode structure similar to infinite chain when the conception of group velocity starts to be applicable.

Since the fabrication and measurement of nanospheres operating in the optical frequency range is difficult, below we demonstrate the proof-of-concept experiment for the microwave frequency range, similar to earlier studies~\cite{Savelev2014_1}. We scale up all the dimensions and perform numerical simulations and experimental studies. We use MgO-TiO$_2$ ceramic disks with permittivity $\varepsilon=16$ and dielectric loss factor of $1{\rm e}^{-3}$. In order to confirm experimentally the frequency dependence of the
Purcell factor, we measure directly the input impedance of an electrically short wire antenna exciting a chain of ceramic cylinders at microwaves [see Fig.~\ref{Fig3_exp}(a)]. We have considered a ceramic chain with the following parameters: the radius of the ceramic cylinder $r=4$~mm, the cylinder height $H=4$~mm, and the period of the chain $a=5$~mm. In such a measurement the Purcell factor is found as a ratio between the real part of input impedance of the antenna in presence of the structure and the real part of input impedance of the same antenna situated in free space~\cite{Krasnok_Purcell_2015}. The radiation resistance of small dipole antennas characterizes the radiated power, and it is equal to the real part of the radiation resistance measured directly in the feeding point. The ceramic cylinders used in the experiment exhibit their individual magnetic dipole resonances at 8.5~GHz. Therefore, to measure correctly the real part of the input impedance of an electrically short antenna (the length is much smaller than the radiation wavelength) in the broad frequency range, we use a monopole over a metal ground plane [a mirror in Fig.~\ref{Fig3_exp}(a)]. Such a setup requires no balanced-to-unbalanced transformer, and it can be measured in a standard way with a single calibrated 50-Ohm coaxial port of vector network analyzer (Rohde\&Schwarz ZVB20). However, the heights of the antenna and the cylinders effectively double due to the metal ground plane. The ceramic cylinders are
arranged in a chain with a thin perforated polyethylene holder, which does not affect the electromagnetic properties of the structure. The monopole antenna is formed by a 0.5~mm--thick core of the coaxial cable going through a hole in a copper sheet. The monopole height above the ground plane is chosen to be as small as 3~mm to avoid its own resonances in the range from 7 to 11~GHz.

The results of numerical simulations are summarized in Fig.~\ref{Fig3_exp}(b). For the Purcell factor calculation in the system of non-spherical particles we use the approach of input impedance extraction~\cite{Krasnok_Purcell_2015}. We observe a strong enhancement of the Purcell factor at frequency of 8.6~GHz up to 65 for the 10-particles chain. In Fig.~\ref{Fig3_exp}(c), we present the results of experimental measurements of the Purcell factor. We observe a good agreement between the experimental data and numerical results. The slight difference between the experimental and theoretical results at the first resonance we explain as follows. The near field of our monopole source is not exactly near field of a dipole (because of the small hole in the metal sheet and of the coaxial cable end). Thus, the conditions of the nearest cylinders excitation in the experiment are different from ones used in the numerical calculations. We notice that for a different orientation of the cylinders in the chain, namely the cylinders axis oriented along the axis of the chain, we observe much smaller values of the Purcell factor.

In conclusion, we have predicted a simple way for achieving a strong Purcell effect for all-dielectric nanostructures without high local-field enhancement. We have shown that Purcell factor can be increased by two orders of magnitude in a finite chain of silicon nanoparticles. Using the eigenmode analysis for an infinite chain, we have demonstrated that the high Purcell factor is associated with the Van Hove singularities. We have confirmed our theoretical predictions by the proof-of-principle microwave experiments with arrays of high-index subwavelength dielectric particles, and have observed 65-fold Purcell factor enhancement for the chain of 10 particles. We believe that the similar effects can be observed for nanoscale structures due to scalability of the all-dielectric photonics.\\

The authors thank E.~A.~Nenasheva and D.~Filonov for a technical help, F.~Capolino and I.~Maksymov for useful discussions. Y.K. thanks E. Yablonovich for highlighting discussions of the interference nature of the Purcell effect and its link to the antenna theory. This work was supported by the Russian Science Foundation (Grant 15-19-30023) and the Australian Research Council.

%\bibliography{References}

%

\end{document}